\documentclass[12pt, a4paper]{scrartcl}

\usepackage[utf8]{inputenc}
\usepackage[T1]{fontenc}
\usepackage[USenglish]{babel}

\usepackage[tracking=true, protrusion=true, expansion=true,final, babel]{microtype}
\usepackage{lmodern}

\usepackage[pdfborderstyle={/S/U/W 1}]{hyperref}

\usepackage[version=4]{mhchem} 

\usepackage{amssymb}

\usepackage{graphicx}
\usepackage[labelfont=bf]{caption}
\usepackage{placeins}

\usepackage[separate-uncertainty]{siunitx}
\DeclareSIUnit\at{at.\%}
\DeclareSIUnit\rpm{rpm}
\DeclareSIUnit\ppm{ppm}

\usepackage[maxbibnames=99, maxcitenames=1, uniquelist=false, style=authoryear-comp, sorting = nyt]{biblatex}
\usepackage{csquotes}

\usepackage{soul} 

\usepackage{authblk}
\usepackage{orcidlink}

\title{Facilitating Atom Probe Tomography of 2D MXene Films by In Situ Sputtering}
\date{}

\author[1,*]{Mathias Krämer \orcidlink{0000-0002-1352-9064}}
\author[2]{Bar Favelukis}
\author[2]{Maxim Sokol}
\author[2]{Brian A. Rosen}
\author[2]{Noam Eliaz}
\author[1,3]{Se-Ho Kim}
\author[1,4,*]{Baptiste Gault \orcidlink{0000-0002-4934-0458}}

\affil[1]{Max-Planck-Institut für Eisenforschung, Max-Planck-Straße 1, 40237 Düsseldorf, Germany}

\affil[2]{Department of Materials Science and Engineering, Tel Aviv University, P.O.B 39040, Ramat Aviv 6997801, Israel}

\affil[3]{Department of Materials Science and Engineering, Korea University, Seoul 02841, Republic of Korea}

\affil[4]{Department of Materials, Royal School of Mines, Imperial College London, London, SW7 2AZ, United Kingdom}

\affil[*]{Corresponding authors: m.kraemer@mpie.de, b.gault@mpie.de}

\begin{document}

\maketitle

\clearpage

\section*{Abstract}

2D materials are emerging as promising nanomaterials for applications in energy storage and catalysis. In the wet chemical synthesis of MXenes, these 2D transition metal carbides and nitrides are terminated with a variety of functional groups, and cations such as \ce{Li+} are often used to intercalate into the structure to obtain exfoliated nanosheets. Given the various elements involved in their synthesis, it is crucial to determine the detailed chemical composition of the final product, in order to better assess and understand the relationships between composition and properties of these materials. To facilitate atom probe tomography analysis of these materials, a revised specimen preparation method is presented in this study. A colloidal \ce{Ti3C2T_z} MXene solution was processed into an additive-free free-standing film and specimens were prepared using a dual beam scanning electron microscope / focused ion beam. To mechanically stabilize the fragile specimens, they were coated using an \textit{in situ} sputtering technique. As various 2D material inks can be processed into such free-standing films, the presented approach is pivotal for enabling atom probe analysis of other 2D materials.

\section*{Keywords}
atom probe tomography, focused ion beam, \textit{in situ} coating, 2D materials, MXenes

\clearpage

\section*{Introduction}

Since the groundbreaking experimental observation of graphene~\parencite{2004Nov}, there has been significant interest in 2D materials and their fascinating functional properties. Following graphene, the successful synthesis of other 2D elemental materials~\parencite{2017Man}, 2D transition metal dichalcogenides~\parencite{2017Manz} and 2D perovskites~\parencite{2019Lan}, to name a few, has significantly expanded the landscape of 2D materials. Because 2D materials have a thickness of one up to a few atomic layers, and thus a high surface-to-volume ratio, they are particularly interesting for applications involving highly surface-active chemical processes, such as energy storage or catalysis~\parencite{2013Nic}. 

MXenes, 2D transition metal carbides and nitrides, have also been the subject of intense research since their discovery by the Barsoum and Gogotsi groups~\parencite{2011Nag}. To date, more than 50 different compositions have been synthesized, not even considering the different and highly variable surface chemistries~\parencite{2021Vah, 2023Ana}. Their top-down synthesis involves two steps, namely the harsh selective etching of the A layer (typically \ce{Al}, \ce{Si} or \ce{Ga}) from a bulk MAX phase(-like) precursor in a typically \ce{HF}-based wet chemical environment, and the subsequent exfoliation of the obtained weakly bonded multilayer MXenes into individual nanosheets by intercalation of cations or molecules~\parencite{2017Alh, 2022Lim}. The general chemical formula of MXenes is written as \ce{M_{n+1}X_nT_z}, characterized by $n+1$ atomic layers of one or more early transition metals M and $n$ interleaved atomic layers of \ce{C} and/or \ce{N}, denoted as X, where \ce{T_z} represents surface termination groups saturating the bare surface during synthesis. Because this synthesis process is scalable~\parencite{2020Shu}, and the MXene nanosheets can be easily processed into free-standing films~\parencite{2020Zha} or printable inks~\parencite{2019Zha}, they are attractive and accessible for industrial applications.

The obtained properties are significantly influenced by the detailed synthesis parameters~\parencite{2023Tha}, which control in particular the flake size, defect density and surface chemistry. A carefully optimized synthesis route can therefore be used to fine-tune the properties of the MXenes, which is comparable to defect engineering in bulk materials~\parencite{2017Li}. However, Shuck pointed out that MXenes are unfortunately often treated as chemicals rather than materials, because the detailed synthesis route is overlooked~\parencite{2023Shu}. For example, in many synthesis protocols for \ce{Ti3C2T_z}, the most studied MXene to date, spontaneous intercalation of \ce{Li} cations~\parencite{2013Luk} is crucial for obtaining large and high-quality, i.e. less defective, monolayer MXene flakes~\parencite{2014Ghi, 2016Lip, 2016San, 2022She}. Although it is known that the presence of \ce{Li} influences the properties of the MXenes~\parencite{2020Che}, it remains a difficult task to localize and quantify \ce{Li} in the material using the commonly applied techniques~\parencite{2021She}.  

Atom Probe Tomography (APT) is an analytical characterization technique with sensitivity to both light and heavy elements and sufficient spatial resolution to address these unanswered questions~\parencite{2021Gau}. Briefly summarized, the atom probe is a time-of-flight mass spectrometer equipped with a position-sensitive detector, where individual atomic or molecular ions are field evaporated from the apex of a sharp, needle-shaped specimen. In combination, the recorded time-of-flight and the impact coordinates of the ions on the detector, allow compositional mapping with a sub-nanometer spatial resolution in 3D \parencite{2020DeG}. Given these capabilities, APT has been increasingly used in recent years to characterize functional nanomaterials, such as nanoparticles for catalysis, as it provides unique compositional insights that enable an understanding of the activity and degradation of these materials~\parencite{2022Li}. 

A critical step towards the APT analysis of nanostructures, including nanoparticles, nanosheets and nanowires, was the development of appropriate specimen preparation approaches, as these materials, unlike bulk materials, do not allow straightforward traditional protocols to prepare needle-shaped specimens~\parencite{2015Fel}. Common workarounds include the deposition of nanoparticles on pre-sharpened needles by electrophoresis~\parencite{2011Ted}, or the fixation of nanoparticles with a micromanipulator and subsequent coating on a support~\parencite{2015Dev} followed by sharpening by focused ion beam (FIB) milling. Attachment on pre-sharpened needles combined with a coating has also been proposed~\parencite{2022Jos}. The deposition of a dense matrix material encapsulating the nanoparticles by atomic layer deposition~\parencite{2015Lar} or metallic electrodeposition~\parencite{2018Kim} enables the utilization of the commonly used specimen preparation using a dual beam scanning electron microscope (SEM)-FIB~\parencite{2007Tho}. In the case of nanowires, their needle-like shape may even allow them to be analyzed without special specimen preparation in an atom probe with a local electrode~\parencite{2006Per, 2013Du}. 

Despite the wide range of compositions available, and the increasing intensity of research on nanostructures, 2D materials such as MXenes have rarely been investigated using APT. Although graphene is occasionally used to coat biological or liquid APT specimens~\parencite{2018Adi, 2020Qiu_a, 2020Qiu_b, 2020Qiu_c}, the analysis of graphene is only incidental. Previous studies have shown how APT can advance our understanding on the detailed composition of 2D materials, such as the incorporation of impurity elements during the wet chemical synthesis of 2D \ce{MoS2}~\parencite{2020Kim}, or the influence of alkali elements from synthesis on the oxidation of \ce{Ti3C2T_z} MXenes~\parencite{2023Krä}. In both cases, the nanosheets were embedded in a metallic matrix, but they are very difficult to localize for targeted sample preparation, and the small number of nanosheets in an average data set limits the statistics. 

Here, APT analysis is performed by taking advantage of the possibility to process MXenes into free-standing films, i.e. a macroscopic stack of nanosheets with the same orientation held together by weak intermolecular forces such as van der Waals forces or hydrogen bonds. APT specimens of a free-standing \ce{Ti3C2T_z} MXene film were prepared by FIB lift-out. Following sharpening, APT specimens were coated \textit{in situ} by ion milling a \ce{Cr} lamella as sputter target~\parencite{2023Woo}, to mechanically stabilize the fragile specimens. In addition to enhanced yield, performance, and increased field-of-view~\parencite{2023Sch}, as well as the reduction of artifacts in the analysis of Li-containing materials~\parencite{2023Sin}, the presented workflow involving the \textit{in situ} coating technique may also be a starting point for a simplified and straightforward APT analysis of 2D materials. 

\section*{Materials and Methods}

\subsection*{Synthesis of \ce{Ti3AlC2} MAX Phase}

\ce{Ti3AlC2} MAX phase was synthesized by solid–liquid reaction. Briefly, \ce{TiC} (\SI{99.5}{\percent}, Alfa Aesar), \ce{Ti} (\SI{99.7}{\percent}, Strem) and \ce{Al} (\SI{99.7}{\percent}, Strem) powders were ball milled at \SI{1800}{\rpm} for a duration of \SI{5}{\minute}. The resulting powder mixture was cold pressed and then heat treated in a furnace at \SI{1500}{\degreeCelsius} for a period of \SI{120}{\minute} under a protective \ce{Ar} environment. Finally, the sintered \ce{Ti3AlC2} MAX phase was ball milled at \SI{1800}{\rpm} for \SI{5}{\minute} to a fine powder, ready for MXene synthesis.

\subsection*{Synthesis of \ce{Ti3C2T_z} MXenes}

For \ce{Ti3C2T_z} MXene synthesis, \SI{0.5}{\g} \ce{LiF} (\SI{99}{\percent}, Strem) was dissolved in \SI{5}{\ml} \SI{10.2}{\mol\per\litre} concentrated \ce{HCl} (\SI{32}{\percent}, Bio-Lab) in a high density polyethylene vial, to prepare the etchant for selective etching of the \ce{Al} layer from the previous sintered \ce{Ti3AlC2} MAX phase. Etching was done by slowly adding \SI{0.5}{\g} of the MAX phase powder to the solution under constant stirring with a magnetic stirrer at \SI{45}{\degreeCelsius} degrees for \SI{24}{\hour}. 

After etching, the complete solution was transferred to a \SI{50}{\ml} centrifuge tube and filled with deionized water (conductivity \SI{0.055}{\micro\siemens\per\centi\metre}). The solution was shaken thoroughly and then centrifuged at \SI{3500}{\rpm} for a duration of \SI{2}{\minute}. The supernatant was decanted, and the remaining sediment was replenished with deionized water, shaken and centrifuged again. Washing was repeated several times until the solution reached a near-neutral pH value, i.e. greater than 6. After washing, the tube containing the sediment was refilled with deionized water and the solution was sonicated in an ice bath for \SI{60}{\minute} to prevent heating. To remove unetched residues of the MAX phase, the solution was centrifuged at \SI{3500}{\rpm} for \SI{30}{\minute} and the black colloidal supernatant containing single layer MXenes was collected. Finally, the colloidal solution with a yield of about \SI{7}{\gram\per\litre} was stored at \SI{5}{\degreeCelsius}.

\subsection*{Preparing Free-Standing \ce{Ti3C2T_z} MXene Film}

\begin{figure}[!htb]
   \centering
   \includegraphics[]{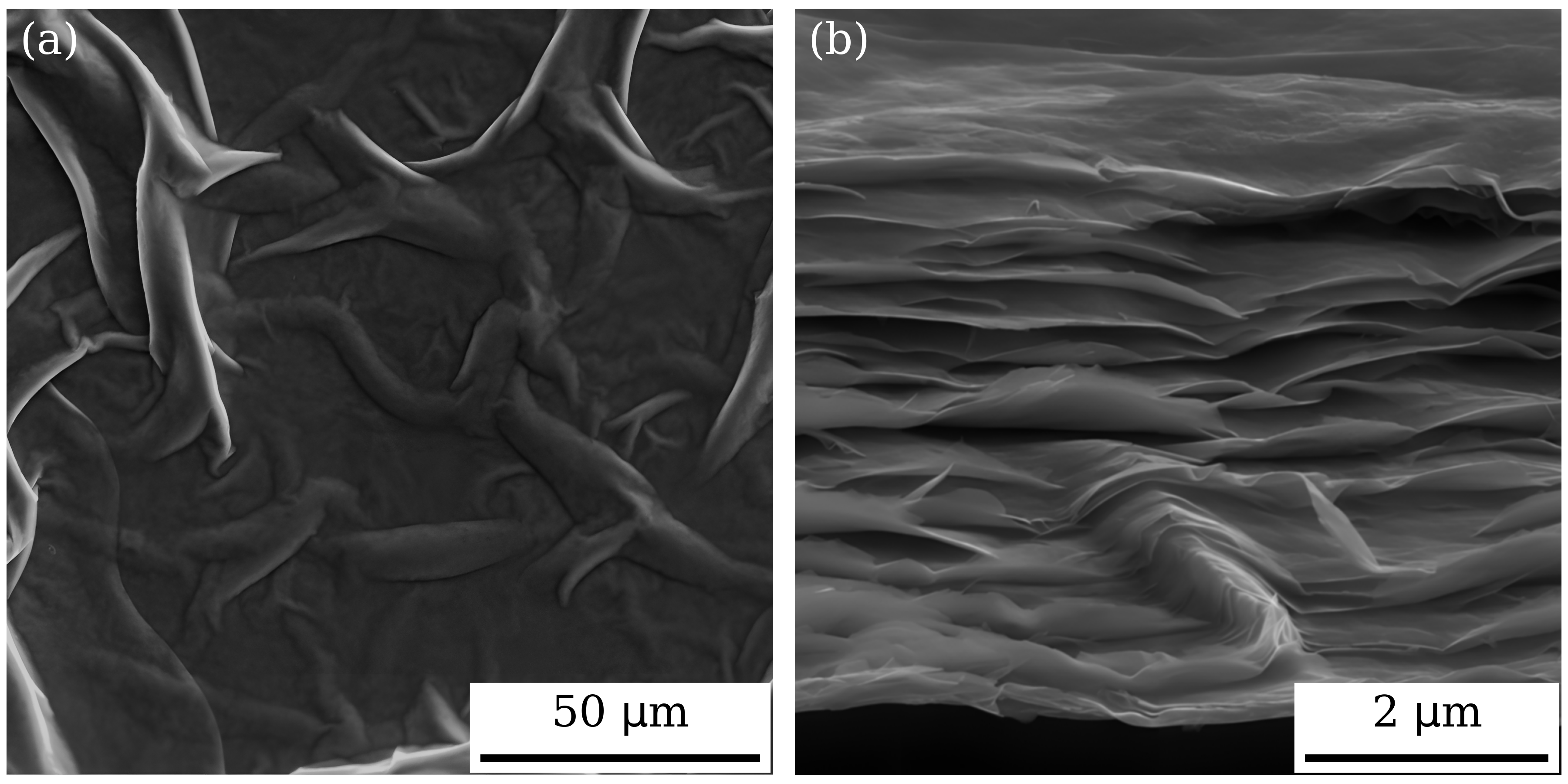}
   \caption{(a) Top and (b) cross-sectional view of the free-standing \ce{Ti3C2T_z} MXene film.}
    \label{fig:FreeStandingFilm}
\end{figure}

The free-standing MXene film was prepared by vacuum filtration. \SI{5}{\milli\litre} of the colloidal solution was poured into a vacuum filtration system using a Celgard\textsuperscript{\textregistered} \num{3501} membrane. After filtration, the film was removed from the membrane. \autoref{fig:FreeStandingFilm} shows both the top (a) and the cross-sectional view (b) of the free-standing film, revealing the horizontal alignment of the nanosheets in the stack.

\subsection*{APT Specimen Preparation}

\begin{figure}[!htb]
   \centering
    \includegraphics[]{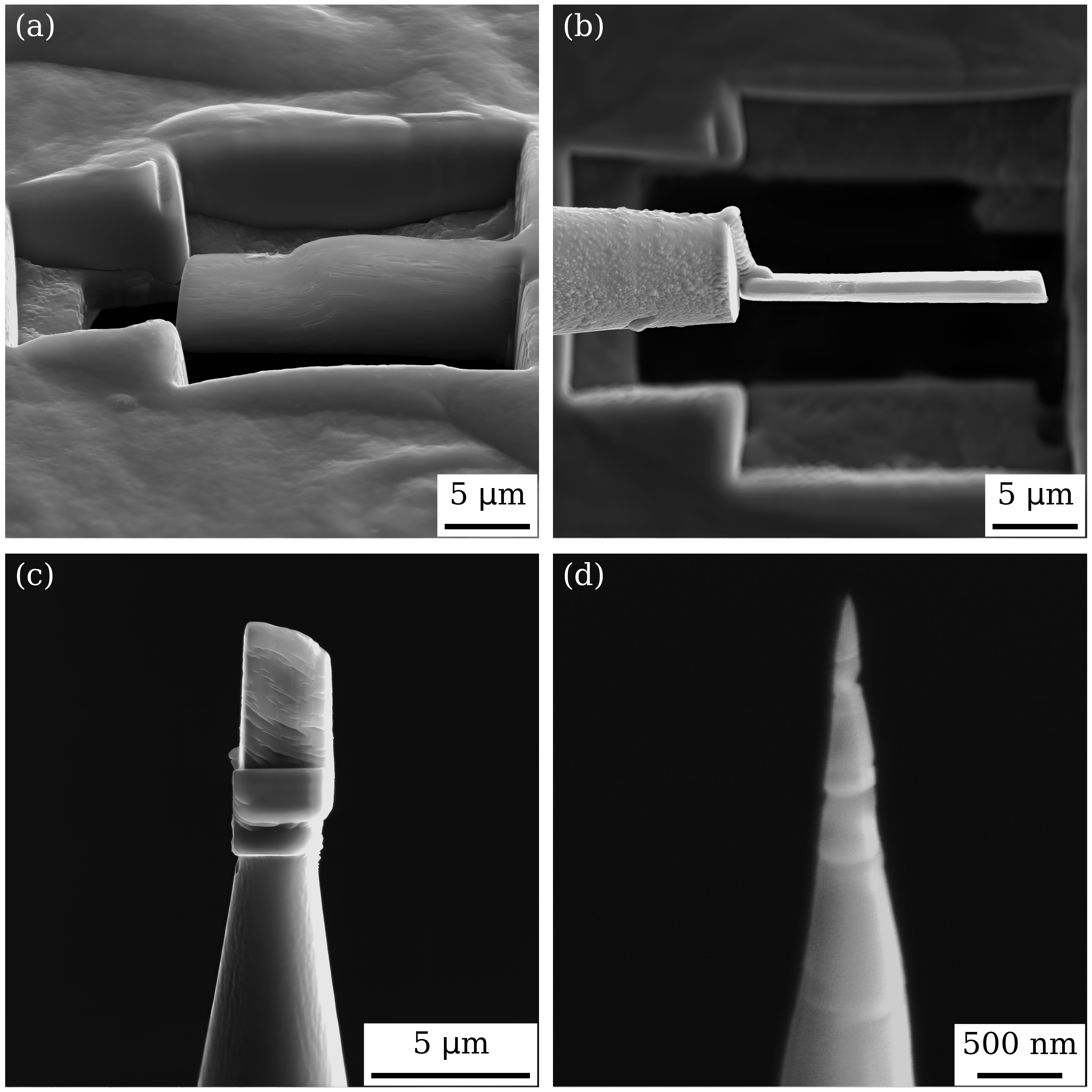}
   \caption{APT specimen preparation of the free-standing \ce{Ti3C2T_z} MXene film following common lift-out and sharpening protocols. (a) Cross-sectional view of a lamella sliced out of the film using the \ce{Ga} ion beam. (b) Lift-out of the lamella, which is attached to a micromanipulator by decomposing a gaseous \ce{Pt/C} precursor from a gas-injection system. (c) Mounted lift-out on a commercial silicon support. (d) Final sharpened APT specimen after annular ion milling. The horizontal stacking orientation of the nanosheets is readily visible.}
    \label{fig:Uncoated_Specimen}
\end{figure}

\begin{figure}[!htb]
   \centering
    \includegraphics[]{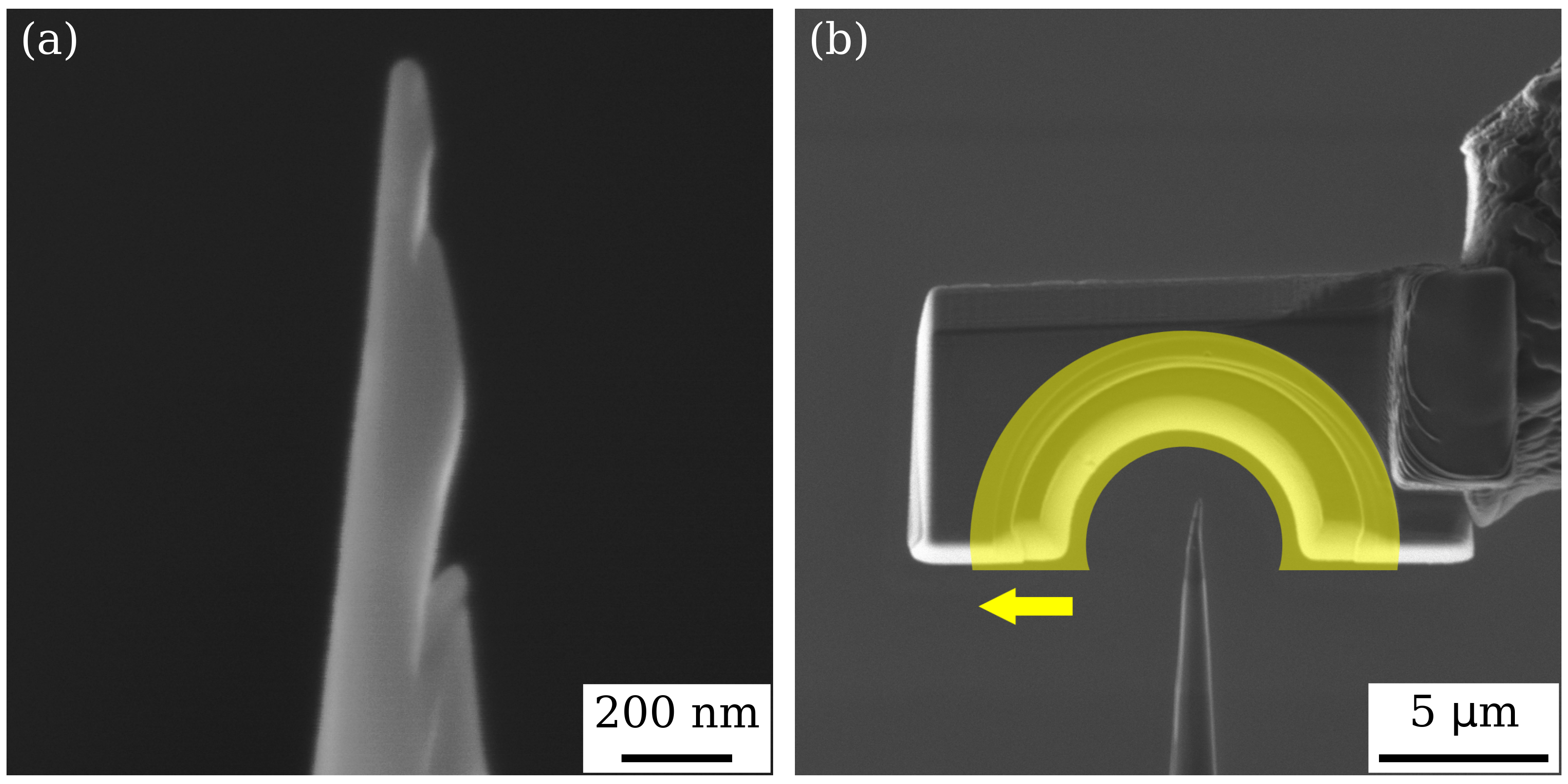}
   \caption{Revised APT specimen preparation for the free-standing \ce{Ti3C2T_z} MXene film utilizing the \textit{in situ} sputtering technique. (a) Uncoated APT specimen. The lift-out was rotated to orient the nanosheets along the specimen. (b) \textit{In situ} coating procedure. The milling pattern is schematically shown in yellow.}
    \label{fig:Coating}
\end{figure}

A first batch of needle-shaped APT specimens was prepared from the free-standing MXene film according to the lift-out and sharpening protocol introduced by~\textcite{2007Tho}, using a dual beam SEM-FIB (Helios Nanolab 600, FEI) equipped with a \ce{Ga} ion source, as illustrated in \autoref{fig:Uncoated_Specimen}~(a) - (d). For a second batch of samples, the lift-out was first attached to a needle able to rotate the lifted-out lamella by \SI{90}{\degree}, thereby aligning the MXene nanosheets more favourably for APT analysis, and then picked up again with the micromanipulator to continue with common specimen preparation. After annular ion milling, specimens such as in \autoref{fig:Coating}~(a) were then additionally coated with \ce{Cr} (\SI{99.9}{\percent}, small leftover of the synthesis of larger ingots from the workshop at the Max-Planck-Institut für Eisenforschung) using an \textit{in situ} sputtering technique described by~\textcite{2023Woo}, building on previous reported works~\parencite{2009Koe, 2023Dou}, and as depicted in \autoref{fig:Coating}~(b). More details on the complete coating workflow can be found in~\textcite{2023Sch}. Sputtering parameters for the ion beam pattern were \SI{30}{\kilo\volt} and \SI{48}{\pico\ampere} for \SIrange{20}{30}{\second}, repeated \num{4} times after rotating the specimen \SI{90}{\degree} each time to ensure a uniform coating. \ce{Cr} as a coating material was chosen in first place for its well-known adhesion properties.

\subsection*{APT Characterization}

APT analyses were performed using either a 5000XS (straight flight path) or a 5000XR (reflectron-fitted) local electrode atom probe (Cameca Instruments), operating in ultraviolet ($\lambda =$ \SI{355}{\nm}) laser-pulsing mode. Parameters were set to a base temperature of \SI{50}{\kelvin}, a laser pulse energy varied between \SI{50}{\pico\joule} and \SI{75}{\pico\joule}, a laser pulsing rate of \SI{125}{\kilo\hertz}, and a target detection rate of \SIrange{5}{10}{} ions per \num{1000} pulses on average. Data reconstruction and analysis was done with AP Suite 6.3 by Cameca Instruments following the default voltage-based reconstruction algorithm.

\section*{Results and Discussion}

\subsection*{Mechanical Instability and \textit{In Situ} Delithiation}

\begin{figure}[!htb]
   \centering
    \includegraphics[width=\textwidth]{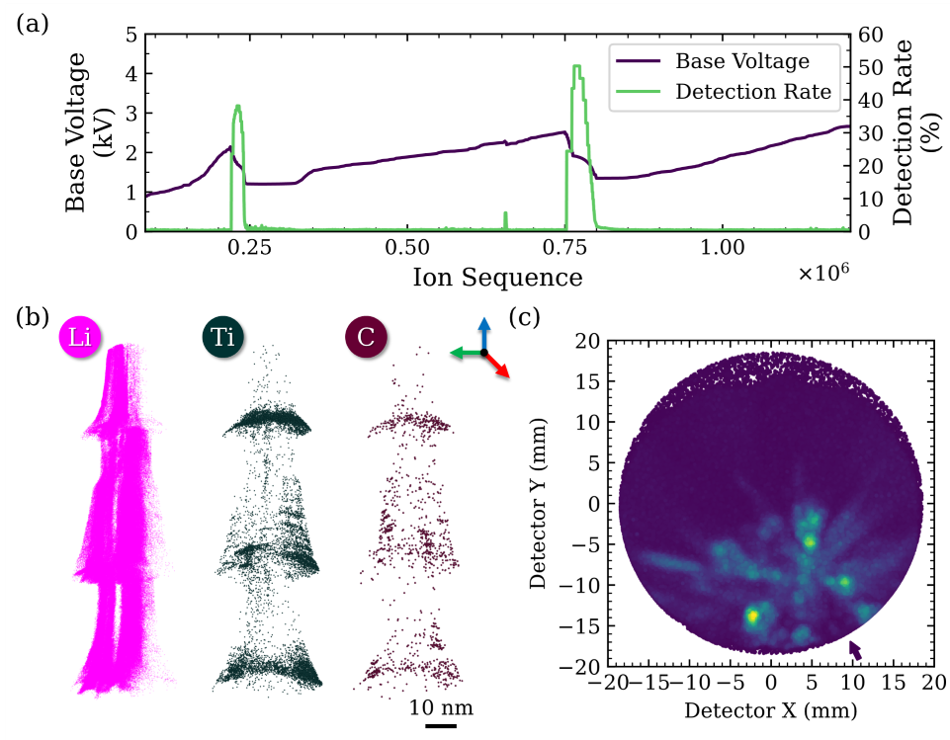}
   \caption{Characteristic APT analysis of the free-standing MXene film prepared following common lift-out and sharpening protocols. (a) Base voltage and detection rate curve. (b) Reconstructed 3D atom maps. (c) Detector hit map. The direction of the laser beam is indicated by an arrow.}
    \label{fig:APT_Uncoated}
\end{figure}

Initial attempts to analyze APT specimens of the free-standing MXene film, prepared as described in \autoref{fig:APT_Uncoated}, faced significant problems with the mechanical stability of the specimens, that may be inherent in the APT analysis of such materials. In this arrangement, the stacking orientation of the nanosheets is perpendicular to the main axis of the APT specimen, as visible in \autoref{fig:Uncoated_Specimen}~(d). Despite the intermolecular interactions between the MXenes nanosheets, they are not strong enough to withstand the high Maxwell stresses arising from the intense electrostatic field applied during the analysis. In addition, there may also be nanovoids between the nanosheets, as visible in the cross-sectional view of the free-standing film in \autoref{fig:FreeStandingFilm}~(b), where intermolecular forces will be almost absent. This results in multiple small fractures, as evidenced by drops in the base voltage curve in \autoref{fig:APT_Uncoated}~(a), which can be explained by a sudden increase in the detection rate due to the detection of several nanosheets breaking off from the specimen apex in very close succession. In almost all cases, measurements are limited to less than a million ions detected before the specimens fractures completely from the silicon support, so the success rate for these samples is rather low. 

\ce{Li} was normally only used in the wet chemical synthesis for exfoliation. However, here up to \SI{90}{\at} \ce{Li} was measured in the analyzed specimens. In the reconstructed 3D atom map in \autoref{fig:APT_Uncoated}~(b), it was observed that at the beginning and after each microfracture, \ce{Li} field evaporates first before other species such as \ce{Ti} are also detected. During APT analysis, the intense electrostatic field can cause preferential migration of \ce{Li} atoms~\parencite{2014Gre}, also known as \textit{in situ} delithiation~\parencite{2017Pfe}, which is a known artifact in the analysis of \ce{Li}-containing materials. Hot spots in the detector map in \autoref{fig:APT_Uncoated}~(c), localized preferentially on the side of the specimen directly illuminated by the laser, are another indicator for this phenomenon, since the temperature rise of the specimen due to the absorbed laser energy increases the mobility and thus the migration of the \ce{Li} atoms~\parencite{2022Kim_a}. Considering the observed mechanical instability of the specimen, the \textit{in situ} delithiation could possibly even favor it, since the sudden deintercalation of \ce{Li} may weaken the intermolecular forces between the nanosheets. 

In summary, it was not possible to perform a reliable APT analysis on specimens of the free-standing 2D material film prepared using the common FIB lift-out and sharpening protocols. On the one hand, the specimens lack mechanical stability, and on the other hand, preferential migration of \ce{Li} atoms prevents detailed compositional analysis. 

\subsection*{\textit{In Situ} Coating for Mechanical Stabilization}

In order to overcome both the mechanical instability and \textit{in situ} delithiation issue that hinder successful APT analysis of the free-standing \ce{Ti3C2T_z} MXene film, specimens were prepared according to the workflow described in \autoref{fig:Coating}. Coating of APT specimens resulted in enhanced yield for a wide range of materials~\parencite{2009Koe, 2013Lar, 2023Sch}, for example by smoothing out the roughness of the specimen surface, but also suppressed \textit{in situ} delithiation in battery materials~\parencite{2022Kim_a, 2023Sin}, or the piezoelectric effect in  perovskite-structured materials~\parencite{2023Kim} by shielding the electrostatic field. By sputtering a material onto the specimen, pores and nanovoids may also be filled, which helps to reduce varying magnification in the reconstruction due to ion trajectory aberrations as well as possible crack tips for premature fracture during analysis~\parencite{2015Pfe}. Various methods have been proposed to eliminate pores in APT specimens, such as electron beam induced deposition~\parencite{2015Pfe, 2020Bar}, electrodeposition~\parencite{2017ElZ, 2017Mou}, liquid metal encapsulation~\parencite{2022Kim_b} or resin impregnation~\parencite{2023Zan}. However, these potential solutions have all their own drawbacks, such as exposing the material to an electrochemical environment, heat, or high pressure, which could alter the chemistry or structure of the material. \textit{In situ} coating, on the other hand, has the advantage that a variety of materials can be used as sputter targets~\parencite{2023Sch}, and it can even be performed at cryogenic temperatures~\parencite{2023Woo}, minimizing the potential impact on the chemistry and structure of the material.

\begin{figure}[!htb]
   \centering
    \includegraphics[width=\textwidth]{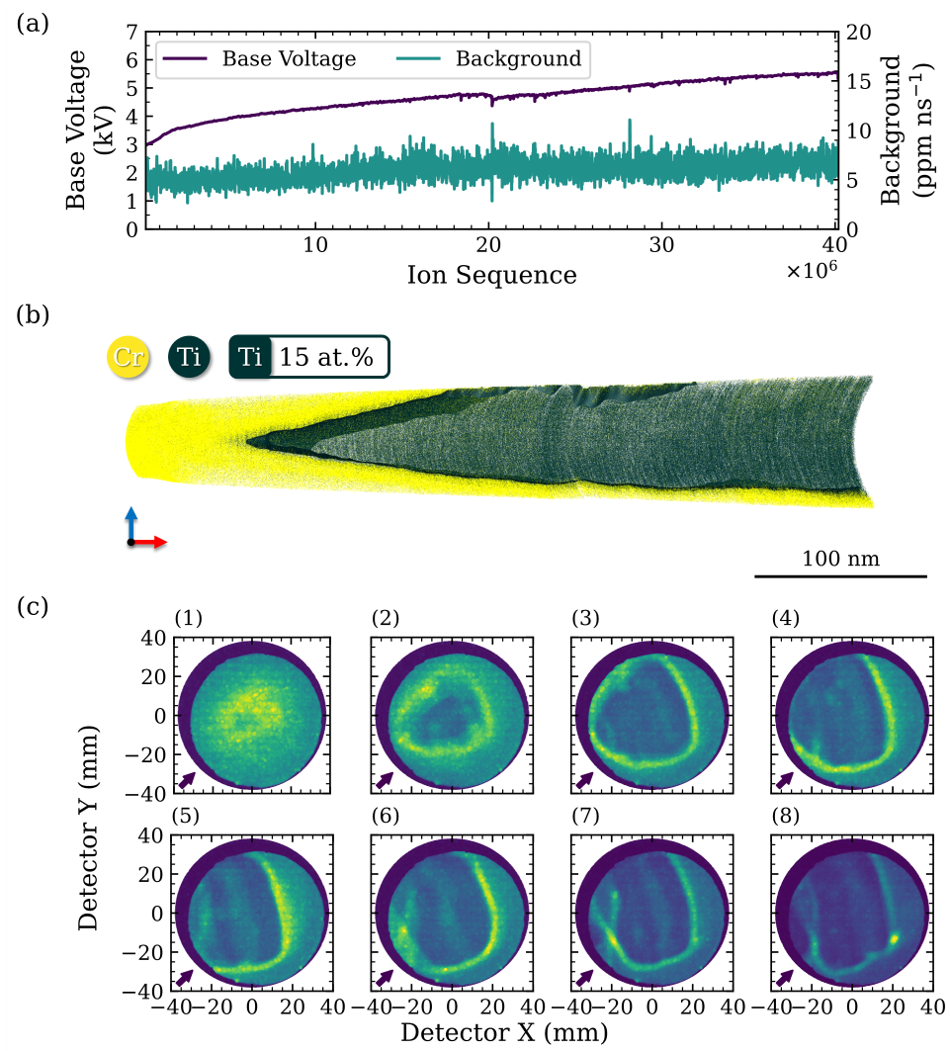}
   \caption{Characteristic APT analysis of the free-standing MXene film prepared following the revised specimen preparation. (a) Base voltage and background level curve. (b) Cross-section of the reconstructed 3D atom map. (c) History of the detector hit map during the measurement, starting at (1). The direction of the laser beam is indicated by an arrow.}
    \label{fig:APT_Coated}
\end{figure}

After metallic coating using the \textit{in situ} sputtering technique, specimens of the free-standing MXene film were successfully analyzed by APT. As visible in the cross-section of the reconstructed 3D atom map in \autoref{fig:APT_Coated}~(b), the as-prepared free-standing MXene film specimen is well coated by \ce{Cr}. Some of the \ce{Cr} coating also seems to have penetrated into the stacked material, and has filled nanovoids between the nanosheets. \autoref{fig:APT_Coated}~(a) provides a plot of the base voltage and the background level curve of these data, both metrics that indicate how stable the measurement was. Except for small drops, the base voltage steadily and smoothly increases, indicating stable field evaporation conditions. The background is largely constant and below \SI{10}{\ppm\per\nano\second} on average. Previous studies have shown that a low and constant background level is essential for the detailed quantification of alkali elements such as \ce{Li}~\parencite{2015San, 2017Lu, 2022Kim_a}, which can be lost in the background from uncorrelated DC field evaporation due to their low expected evaporation field compared to the other elements composing the material~\parencite{1978Tso}.   

Compared to previous work, where exfoliated \ce{Ti3C2T_z} MXenes were electrodeposited into \ce{Co}~\parencite{2023Krä}, significantly larger volumes of the material of interest were measured. To illustrate the improvement, the detected \ce{Ti} ion counts, including the contribution of decomposed molecular ions, from the APT data of the exfoliated MXenes from \textcite{2023Krä} and the free-standing MXenes films were compared. While less than \num{0.1} million \ce{Ti} ions were detected in the data set for the exfoliated MXenes embedded in \ce{Co}, between \num{1.5} and \num{2} million \ce{Ti} ions were collected in the free-standing MXene film in different measurements. All these values refer to data acquired on a 5000XR instrument with a detection efficiency of \SI{0.52}{\percent}. Notwithstanding the much simpler and time-saving specimen preparation workflow, the APT data obtained for the nanomaterial is significantly larger and therefore statistically more reliable. In addition, the detector hit maps of the \ce{Cr} coated samples in \autoref{fig:APT_Coated}~(c) do not show characteristic hot spots, that would indicate a preferential migration of \ce{Li} atoms. This confirms that the \ce{Cr} coating prevents the \textit{in situ} delithiation, as was suggested in previous studies~\parencite{2023Sin}. 

Besides \ce{Cr}, other materials may also be considered as sputter targets in the future. Although it has been shown that specimens can be easily coated with \ce{Cr} using the \textit{in situ} sputtering technique, it has some disadvantages for this particular case. During the sputtering process, the clean \ce{Cr} layer is constantly passivated with an oxide layer~\parencite{2023Sch}, despite the vacuum inside the SEM-FIB. Since the coating also penetrates into the free-standing film specimen itself to fill nanovoids, the detailed determination of the oxygen content of the nanomaterial becomes nearly impossible. Therefore, \ce{Cr} should be replaced as coating material by a more chemically inert metal in further studies, depending on the adhesion and the probability of peak overlap with the material of interest in the APT mass spectrum.

\section*{Conclusion}

Despite the great interest in understanding the functional properties of nanomaterials by characterizing their detailed composition, 2D materials have rarely been studied using APT. Using \ce{Ti3C2T_z} MXenes as an example, the common SEM-FIB specimen preparation has been revised utilizing an \textit{in situ} sputtering technique to facilitate the APT analysis of 2D materials. By processing the colloidal MXene solution into an additive-free free-standing film to enable a common lift-out and sharpening procedure, and by coating the APT specimen, it was possible to acquire relatively large volumes of the 2D material with high data quality. The coating stabilizes the fragile specimen, but also prevents the \textit{in situ} delithiation of Li, which was incorporated into the material during synthesis. As materials other than MXenes can also be processed into free-standing films~\parencite{2007Dik, 2023Li}, the presented workflow has the potential to be a starting point to study the detailed composition of 2D materials with APT.

\FloatBarrier


\section*{Acknowledgment}

The authors acknowledge financial support from the German Research Foundation (DFG) through DIP Project No. 450800666. The support to the FIB and APT facilities at MPIE by Uwe Tezins, Andreas Sturm and Christian Broß is gratefully acknowledged.


\section*{Conflict of Interest}

The authors declare no conflict of interest.


\section*{Data Availability Statement}

The data that support the findings of this study are available from the corresponding authors upon reasonable request.


\printbibliography


\end{document}